\documentclass[superscriptaddress,prx]{revtex4-2}
\usepackage[utf8]{inputenc}

\usepackage{bm}
\usepackage[dvipsnames]{xcolor}
\usepackage{amsmath}
\usepackage{amssymb}
\usepackage{graphicx}
\newcommand{\ii}{\mathrm{i}}
\newcommand{\dd}{\mathrm{d}}

\begin{document}

\title{Many-defect solutions in planar nematics:  interactions, spiral textures and boundary conditions}
%keywords:
%planar nematic, liquid crystal, topological defects, conformal mapping

\author{Simon \v Copar}
\email{simon.copar@fmf.uni-lj.si}
\affiliation{
 Faculty of Mathematics and Physics, University of Ljubljana, Ljubljana, Slovenia
}%
\author{Žiga Kos}
\email{ziga.kos@fmf.uni-lj.si}
\affiliation{
 Faculty of Mathematics and Physics, University of Ljubljana, Ljubljana, Slovenia
}
\affiliation{
Department of Condensed Matter Physics, Jožef Stefan Institute, Ljubljana, Slovenia
}
\affiliation{International Institute for Sustainability with Knotted Chiral Meta Matter (WPI-SKCM$^2$), Hiroshima University, Higashi-Hiroshima, Japan}

\begin{abstract}
From incompressible flows to electrostatics, harmonic functions can provide solutions to many two-dimensional problems and, similarly, the director field of a planar nematic can be determined using complex analysis. We derive a closed-form solution for a quasi-steady state director field induced by an arbitrarily large set of point defects and circular inclusions with or without fixed rotational degrees of freedom, and compute the forces and torques acting on each defect or inclusion. 
We show that a complete solution must include two types of singularities, generating a defect winding number and its spiral texture, which have a direct effect on defect equilibrium textures and their dynamics.
The solution accounts for discrete degeneracy of topologically distinct free energy minima which can be obtained by defect braiding. 
The derived formalism can be readily applied to equilibrium and slowly evolving nematic textures for active or passive fluids with multiple defects present within the orientational order.

\end{abstract}

\maketitle

\section{Introduction}

Interactions, dynamics, and orientation of topological defects~\cite{alexander2012colloquium} govern the behaviour of materials with liquid crystal-like orientational order. 
Equilibrium multi-defect textures can be realised by photo patterning of the alignment axis on one of the surfaces~\cite{meng2023topological} or by patterning of micro-pillars~\cite{kim2018mosaics}. Such photo-patterned templates can guide the swimming direction of bacteria~\cite{peng2016command,mushenheim2014dynamic} and artificial swimmers~\cite{hernandez2014reconfigurable}, produce a tuneable photonic response~\cite{kim2022quasicrystalline}, or can be used to study elastic properties and reconfigurations of nematic disclinations~\cite{nys2023controlled,jiang2022active,jiang2023designing,yi2023line}. 
Topological defects play a major role also for other symmetries of the orientational order parameter, for example, 
escaped structure in nematic cells can exhibit defects in the effective two-dimensional polar vector field~\cite{pieranski2021physics}, polar order can emerge in ferroelectric nematics~\cite{sebastian2023polarization} and in active biological materials~\cite{prost2015active}. Disclinations in the hexatic phase and their mutual interactions are a key mechanism in the melting of two-dimensional crystals~\cite{halperin1978theory}.
Distinctly out of equilibrium, driven nematic layers~\cite{mur2022continuous} and active nematics~\cite{shankar2022topological} are characterized by the proliferation and annihilation of topological defects.
In active nematics, defects show disordered dynamics, called active turbulence, or form ordered textures~\cite{decamp2015orientational}, and have biologically-relevant properties~\cite{saw2017topological}. 
One approach towards modelling driven or active defect dynamics are particle-like models~\cite{keber2014topology,brown2020theoretical,zhang2020dynamics,giomi2014defect}, based on the description of nematic alignment, where the director field is in equilibrium, apart from the positions and orientations of defects. Interactions and self-propagation of defects lead to their movement and reorientations.
As shown by the above examples, analytical models of various liquid crystalline textures are needed as initial or boundary conditions and, more importantly, to derive general theorems about emergent orientational structures and their dynamics. 

Complex analysis is an important tool to generate solutions of field theories in two dimensions.
As all complex analytical functions $\Phi$ solve the Laplace equation, models with
energy proportional to $(\nabla \Phi)^2$ can be solved by finding analytical functions
that satisfy the boundary conditions. Additionally, conformal mapping can be used
to transform the problem to a domain where solution can be readily obtained.
A well known example is 2D hydrodynamics of ideal incompressible fluids, with Kutta-Joukowsky
theory giving the lift generated by an airfoil by transforming a solution on a unit circle to a realistic airfoil domain~\cite{pozrikidis2011introduction}. 
Complex analysis is also often used to solve for two-dimensional potentials in electrostatics~\cite{feynman1989volume}.
A similar approach can be taken for fluids with an orientational order parameter isomorphic to $e^{\ii n\phi}$, where $n$ describes the polar ($n=1$), nematic ($n=2$), or higher-order symmetry, and $\phi$ the polar angle of the orientational order. For example, alignment of the nematic liquid crystals restricted to the 2D Euclidean plane can be described by a free energy $\frac{K}{2}\int (\nabla \phi)^2 \dd^2 r$, where $K$ is the elastic constant and $\vec{n}=(\cos\phi,\sin\phi)$ is the director field. The equilibrium texture is solved by harmonic functions $\phi$ and singular points of $\phi$ correspond to topological defects (disclinations).

The conformal description of topological defects is regularly used
to generate analytical models of liquid crystalline textures.
Solutions for non-trivial planar geometries are often achieved by conformal mapping \cite{DavidsonAJ_EurJApplMath23_2012,vanBijnenRMW_PhysRevE86_2012,TarnavskyyO_LiqCryst45_2018,TarnavskyyOS_LiqCryst47_2020,TarnavskyyOS_LiqCryst,chandler2023nematic}, or by stereographic projection for spherical domains \cite{VitelliV_PhysRevE74_2006,BrownAT_SoftMatter16_2020}, for describing defects dynamics in active nematics in Q-tensor formalism \cite{VafaF_arXiv}, and active nematic textures in the director formalism \cite{houston2023colloids}. Complex field approach can also be used to solve for a shape of nematic domains with a free boundary under different anchoring conditions \cite{RudnickJ_PhysRevLett74_1995,vanBijnenRMW_PhysRevE86_2012}.
Some of the well known solutions are the core energy of a single nematic defect \cite{deGennesPG_1993,KlemanM}, the director field and the interaction energy of a pair of defects \cite{TangX_SoftMatter13_2017}, and drag force and mobility tensor for moving defects~\cite{KlemanM,TangX_SoftMatter15_2019}.
These solutions focus on topological defects, where the director field rotates by a multiple of $\pi$ along a loop circumnavigating the defect core.
Another singular solution of the planar Laplace equation is a type of defect, where the director polar angle changes logarithmically with the distance from the defect core~\cite{KlemanM}. Such twisted solutions with a radially varying director 
naturally occur in, for example, `magic' spiral problems~\cite{deGennesPG_1993,WilliamsDRM_PhysRevE50_1994,LewisAH_StudiesinAppliedMathematics138_2017}.
However, a solution for an arbitrary number of defects with a given orientation and boundary condition has to include a combination of topological defects and logarithmic singularities. No such general framework has yet been formulated.

In this paper, we develop a complete formalism for planar nematic textures involving an arbitrarily large set of point defects with arbitrary winding numbers. We introduce a complex winding number (spiral charge) to take into account
twisted spiral contributions to the director around the defect, and derive the conditions required to
obtain the optimal spiral charges when rotational states of some defects are constrained.
We discuss the relationship between topological invariants of director textures and branch cuts of the complex representation of the field. Finally, we derive a general expression for forces and torques on topological defects and show how spiral charges lead to non-central forces, spiral annihilation trajectories and defect braiding.

\section{Conformal formulation of the elastic free energy}

To obtain a conformal description of the director field and the free energy, we express positional coordinate pairs in the form of a complex number, $z=x+\ii y=r e^{\ii\theta}$. We can express the director field in the complex notation as $\bm{n}=\cos\phi+\ii \sin\phi=e^{\ii\phi}$, where $\phi$ is the director tilt angle. Local equilibrium corresponds to $\nabla^2\phi=0$, which we want to satisfy by writing $\phi$ as an analytic function. In order to do so, we take $\phi$ as a real part of a complex-valued function  
\begin{equation}
    \Phi(z)=\phi(z)+\ii \psi(z)
\end{equation}
and define the director tilt angle as $\phi=\mathrm{Re}(\Phi)$.
While $\psi=\mathrm{Im}(\Phi)$ has no direct physical meaning, it is connected to $\phi$ through Cauchy-Riemann equations for analytic expressions of $\Phi$, which as we show in Eq.~\ref{eq:F} considerably simplifies the evaluations of the free energy integrals.
Please note that instead of directly writing an analytic expression for $\Phi(z)$, we can deduce it from the construction of the director field as $\bm{n}=w(z)/|w(z)|$ by introducing
\begin{equation}
    w(z)=e^{\ii \Phi(z)}=e^{-\psi(z)} e^{\ii \phi(z)}.
\end{equation}
The benefit of this approach is that the poles and zeroes of the function $w(z)$ correspond to the defects in the nematic, so we can ensure defect positions and winding numbers by writing $w(z)$ in a factorized form (see Section~\ref{sec:spiral}).
$\Phi (z)$ can be directly computed from the expression for $w(z)$ by taking its complex logarithm $\Phi(z)=-\ii\log (w(z))$. Due to the nature of the complex logarithm, the branch cuts in $\Phi (z)$ are unavoidable and must be carefully considered when evaluating the free energy.

The nematic elastic energy can be expressed in the regime of one-elastic constant $K$ as
\begin{equation}
F=\frac{K}{2}\iint |\nabla\phi|^2\dd x\,\dd y.
\end{equation}
From now on we will omit the constant pre-factor $\tfrac{K}{2}$. We can reparameterize the free energy by the use of Green's identity 
\begin{equation}
F = \oint \phi(\nabla\phi\cdot\bm{\nu})\dd l=\oint \phi(\nabla\psi\cdot \bm{t})\dd l=\oint \phi\frac{\partial\psi}{\partial l}\dd l=\oint \phi\,d\psi,
\end{equation}
where the integration is performed over the counterclockwise contour, parameterized by the arc length $l$, an outward normal $\bm{\nu}$, and a tangent $\bf t$. We have considered the local equilibrium condition $\nabla^2\phi=0$ and the relation $\nabla\phi\cdot \bm{\nu}=\nabla\psi\cdot\bm{t}$, which is obtained from Cauchy-Riemann equations for $\Phi=\phi+\ii \psi$.
Considering $|\nabla \phi|^2 = |\nabla\psi|^2$ which follows from Cauchy-Riemann equations, and exchanging the roles of $\psi$ and $\phi$, we can also derive
\begin{equation}
F=\oint \phi\,d\psi=-\oint \psi\,d\phi.
\label{eq:F}
\end{equation}
With appropriate choices of contours, and taking into account branch cuts, Eq.~\ref{eq:F} is easy to compute, as we show in Section~\ref{sec:pair}. For instance, along the curves of constant $\psi$ or $\phi$, the integrals in Eq.~\ref{eq:F} simply reduce to differences.

\section{Spiral charge}
\label{sec:spiral}
\begin{figure}
    \centering
    \includegraphics[width=\textwidth]{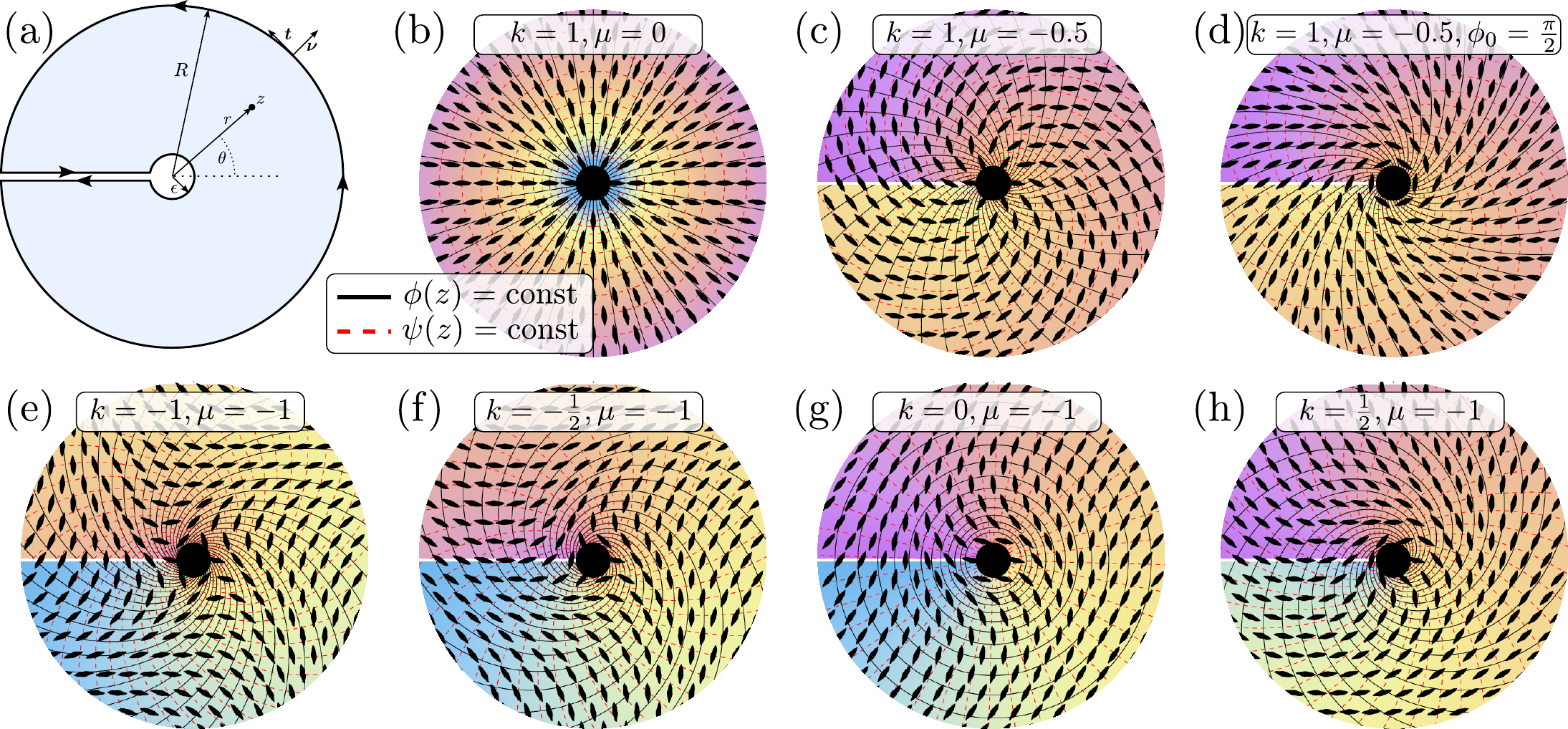}
    \caption{Point defects with different complex winding numbers. (a) A schematic showing the integration contour avoiding the  branch cut, and the notation for the polar coordinates, and the radii of the outer and inner contour. (b-c) A radial hedgehog without and with spiral charge. (d) has an additional $+\pi/2$ rotation to induce a tangential anchoring on the central defect. (e-h) Defects of other winding numbers with added spiral charge. The colour shading shows $\psi(z)$, which has a branch cut on the negative real axis if $\mu\neq 0$. The more familiar branch cut in the $\phi(z)$, which is present at all $k\neq 0$, is not shown. Contours of $\phi$ and $\psi$ form an orthogonal curvilinear grid obeying the conformal properties.}
    \label{fig:spiral}
\end{figure}

We construct point defect solutions as complex functions with a singularity  at the coordinate origin in the form of
\begin{equation}
w(z)=e^{\ii \phi_0}(z/\epsilon)^{k+\ii\mu}=(r/\epsilon)^{k+\ii\mu}e^{\ii \theta(k+\ii \mu)+\ii \phi_0},
\label{eq:power_expression}
\end{equation}
where $k$ and $\mu$ are the winding number and the spiral charge, respectively, which we discuss in detail below. To satisfy the nematic symmetry $\mathbf{n}\sim-\mathbf{n}$, the winding number can take any half-integer value. The denominator $\epsilon$ ensures correct dimensions and can be set to the size of the singular defect core, or the size of a particle in case the defect is bound by a larger circular inclusion.
When $\mu=0$, Eq.~\ref{eq:power_expression} leads to the well known ansatz for 2D nematics, expressed as square roots of complex rational functions with poles and zeroes corresponding to the defects with positive and negative winding numbers. The additional rotation $\phi_0$ corresponds to a rigid rotation of the defects around the center for $k\neq 1$ and to the parametric family of radial and circular hedgehogs for $k=1$ (Fig.~\ref{fig:spiral}b-d).

We can extract the complex polar angle $\Phi(z)=\phi(z)+\ii \psi(z)$ by taking a logarithm of Eq.~\ref{eq:power_expression}:
\begin{align}
    %\Phi(z)=&\phi_0 -\ii (k+\ii \mu) \log \frac{z}{\epsilon}=\phi_0 +(k+\ii \mu)(\theta-\ii\log \frac{r}{\epsilon})\\
    \phi(z)=&\phi_0+k\theta+\mu \log \frac{r}{\epsilon},\label{eq:phi}\\
    \psi(z)=&\mu \theta-k\log \frac{r}{\epsilon}\label{eq:psi}.
\end{align}
The director polar angle $\phi$ matches the standard $\mu=0$ defect profile at $r=\epsilon$, and twists outwards logarithmically (Fig.~\ref{fig:spiral}b-h). Unlike the half-integer restricted winding number $k$, the spiral charge $\mu$ can assume any real value, representing the pitch of a logarithmic spiral seen in the contours of constant angle $\phi(z)$ (Fig.~\ref{fig:spiral}). This radius-dependent twisting is often disregarded in literature unless the geometry explicitly requires it \cite{deGennesPG_1993,KlemanM}.
The continuous nature of $\mu$ ensures this quantity is not topologically protected from changing. It can relax to reduce the overall elastic free energy of the system.

The singularity at the origin has a diverging elastic energy, the nematic region is truncated at a finite radius $\epsilon$, representing the size of the defect core. In the far field, we bound the system at a large radius $R$, and integrate over the contour in Figure~\ref{fig:spiral}a. The elastic energy
\begin{equation}
    F=2\pi (k^2+\mu^2)\ln\frac{R}{\epsilon}
    \label{eq:energy_single}
\end{equation}
generalizes the well known result at $\mu=0$ \cite{KlemanM} and diverges at  $R\to \infty$.  Both the half-integer winding number $k$ and the spiral charge $\mu$ can be combined into a complex generalisation of the winding number $\kappa = k+\ii \mu$, which is a topological invariant in the sense that the quantity
\begin{equation}
    \kappa = k + \ii \mu = \oint\frac{\partial \Phi}{\partial l} \dd l
\end{equation}
equals the sum of the complex winding numbers of the defects enclosed by the integration path. Its nonzero value reflects that $\Phi$ is not an exact form, and points to the existence of branch cuts crossing the circuit around the defect.
In the far field, the energy (Eq.~\ref{eq:energy_single}) will diverge unless the total $\kappa$ equals zero. As such, we expect that the principle of topological charge neutrality in bulk samples extends to the complex winding numbers. In constrained samples, the total real winding number $k$ can be fixed by the boundary conditions, and the total imaginary winding number $\mu$ can equilibrate at a nonzero value according to free energy minimization. The boundary conditions also set the uniform phase $\phi_0$.

\section{Pair interactions}
\label{sec:pair}

\begin{figure}
    \centering
    \includegraphics[width=\textwidth]{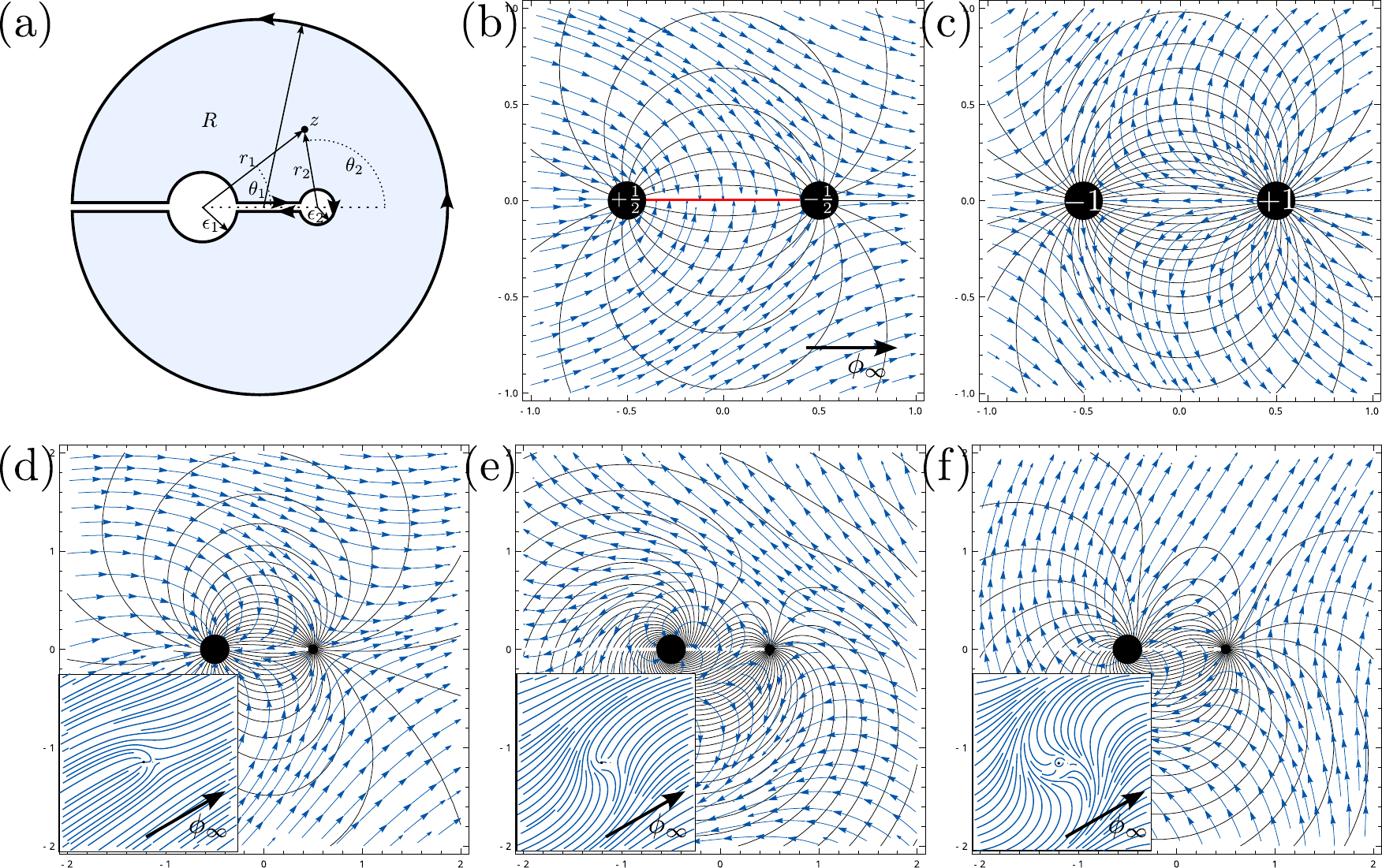}
    \caption{Defect pairs with director field streamlines (blue) and contours of a constant orientation $\phi$ (black). (a) A sketch of the integration contour and the notation for a pair of defects with in principal different core radii. (b,c) well known solutions for a pair of $\pm \tfrac12$ and $\pm 1$ defects with zero spiral charge terms. The case of $\pm \tfrac12$ shows a sign discontinuity in the streamlines due to half-integer winding numbers (shown in red). (d-f) Three solutions for a pair of fixed position defects with prescribed far field direction $\phi_\infty$, prescribed homeotropic anchoring on the radial defect (left), and unconstrained hyperbolic defect (right). The solutions differ by an increasing number of $n\pi$ windings on the radial defect. The (d) configuration has the lowest energy, but all are local minima.}
    \label{fig:dimer}
\end{figure}

Consider now a pair of topological defects with winding numbers $\kappa_1=k_1+\ii \mu_1$ and $\kappa_2=k_2+\ii \mu_2$ at positions $z=\pm  d/2$. 
We allow the core radii to be different, which is reasonable if the winding numbers themselves differ, or if one of the defects is induced by the anchoring on a small circular colloidal inclusion (see Fig.~\ref{fig:extra-winding} for an example). 
The complex director angle can be written as a direct sum of two defect profiles:
\begin{equation}
    \Phi(z)=\phi_0 - \ii \kappa_1 \log \frac{z+d/2}{\epsilon_1}- \ii \kappa_2 \log \frac{z-d/2}{\epsilon_2}
\end{equation}
and split into components,
\begin{align}
    \phi=&\phi_0 + k_1 \theta_1+\mu_1 \ln \frac{r_1}{\epsilon_1} + k_2 \theta_2+\mu_2 \ln \frac{r_2}{\epsilon_2},\\
    \psi=&\mu_1 \theta_1 - k_1 \ln \frac{r_1}{\epsilon_1}+\mu_2 \theta_2 - k_2 \ln \frac{r_2}{\epsilon_2}.
\end{align}
The relative polar positions $(r_1,\theta_1)$ and $(r_2,\theta_2)$ from each defect to the point in question (see Fig.~\ref{fig:dimer}a for a schematic) are a useful representation, because the $\theta_{1,2}$ parts control the position of the branch cuts by deciding where along the circle we make the jump by $2\pi$. If the sum of the winding numbers is zero, $\kappa_1+\kappa_2=0$, the branch cut can be localized to the line between the defects.

The defect pair free energy is given by integrating Eq.~\ref{eq:F} along the contour shown in Fig.~\ref{fig:dimer}a. Considering the confinement radius $R$ that is much larger than defect-defect separation $d$, yields the free energy that completely decouples the winding numbers and the spiral charges, $F=F_k+F_\mu$:
\begin{align}
F_k&=2\pi \left\lbrace (k_1+k_2) \left[k_1 \log \frac{R}{\epsilon_1}+ k_2 \log \frac{R}{\epsilon_2}\right]-k_1k_2 \left[\log \frac{d}{\epsilon_1}+\log \frac{d}{\epsilon_2}\right] \right\rbrace\\
F_\mu &=2\pi \left\lbrace (\mu_1+\mu_2) \left[\mu_1 \log \frac{R}{\epsilon_1}+ \mu_2 \log \frac{R}{\epsilon_2}\right]-\mu_1\mu_2 \left[\log \frac{d}{\epsilon_1}+\log \frac{d}{\epsilon_2}\right] \right\rbrace.
\end{align}
The first term in each part describes the far-field contribution which is positive definite, and vanishes only when the system is charge neutral. This is fulfilled only when the winding numbers are opposite, and when the defects are spiralling in oppositely oriented but equally tight spirals. The second terms describe the pair interaction proportional to the coupling of the winding numbers, and the coupling of the spiral charges, which act as two independent charges that each defect possesses -- one quantized and one continuous, with opposite charges feeling an attractive force.

The free energy is the basis for finding equilibrium director configurations at fixed boundaries, or computing forces and torques acting on defects and inclusions.
The complete symmetry between winding and spiral charges with no cross terms indicates that the only way for these contributions to couple is through boundary conditions, which we will explore later.

\section{Energy minimization}
Consider a problem of finding the director solution for fixed boundaries, which corresponds to a fixed far-field and fixed defect positions and orientations. As the winding numbers are topologically protected, the remaining degrees of freedom of the system are the global phase $\phi_0$ and spiral charges $\mu_i$. These degrees of freedom are not coupled by the elastic free energy itself, but additional constraints may impose relations between them. Boundary conditions are requirements imposed on the orientation $\phi(z)$ at the outer edge or on the defect surfaces, when they are realized as  particle inclusions or fixed by external fields. To apply the boundary conditions, $\phi(z)$ must be well defined on the boundary, which requires correct handling of branch cuts. In this section, we show energy minimization for a defect pair and a fixed far-field, while in the next section we provide a general formalism for an arbitrary number of defects.

The free energy is a positive definite quadratic form in terms of the spiral charges $\mu_1$ and $\mu_2$, and in the absence of boundary conditions, reaches a global minimum when all spiral charges are zero. Applying constraints changes that. Consider an example with a $k_1=+1$ defect enforced by a circular inclusion with a radius $\epsilon_1$ enforcing radial director, an accompanying $k_2=-1$ defect  with core radius $\epsilon_2$ pinned at the distance $d$ along the $x$ axis, and homogeneous far-field in the direction $\phi(R)=\phi_\infty$.

The boundary condition on the far-field director (Eq.~\ref{eq:phi} under the limit $R\gg d$) yields the following constraint:
\begin{equation}
\phi(R)\approx \phi_0 + \mu_1 \log \frac{R}{\epsilon_1}+\mu_2 \log \frac{R}{\epsilon_2}=\phi_\infty.
\label{eq:constraint1}
\end{equation}
To apply the boundary condition on the radial particle, we decompose the director into the contribution affected by the branch cut, and the rest:
\begin{equation}
\phi = \underbrace{k_1 \theta_1 + k_2 \theta_2}_{\phi_k} + \underbrace{\phi_0 + \mu_1 \ln \frac{r_1}{\epsilon_1}+ \mu_2 \ln \frac{r_2}{\epsilon_2} }_{\phi_\mu}
.
\label{eq:phi_components}
\end{equation}
Let both angles $\theta_1$ and $\theta_2$ have an origin pointing to the right, as shown in figure \ref{fig:dimer}a, and range in the interval $(-\pi,\pi)$. On the path around the radial defect at $r_1=\epsilon_1$, the part $\phi_k(\theta_1)$ has a value of $\pi+\theta_1$ on the upper half-space, and $-\pi+\theta_1$ on the lower half-space. This is consistent with the radial condition on the particle, but it would also be consistent if any integer multiple of $\pi$ were added or if the branch cuts were chosen in a different way. Note that $\phi_k$ has a constant rate with ${\rm d}\phi_k/{\rm d}\theta_1=k_1$ only in the limit $\epsilon_1\to 0$, so the boundary condition is not met \emph{exactly} for finitely sized particles. Similarly, the $\phi_\mu $ is not constant on this path in general, but in approximation, we can take the value at the center of the defect:
\begin{equation}
\phi_\mu(z_1) \approx \phi_0 + \mu_2 \ln \frac{d}{\epsilon_2} = n\pi.
\label{eq:constraint2}
\end{equation}
This approximation is valid for $d\gg \epsilon_1$, and is useful for our discussion, as it will allow us to draw more general conclusions. 
As the free energy is a quadratic form, and all the constraints are linear in $\phi_0$, $\mu_1$ and $\mu_{2}$, this is an exactly solvable problem. With two constraints (Eqs.~\ref{eq:constraint1} and \ref{eq:constraint2}) and three variables ($\phi_0$, $\mu_1$ and $\mu_{2}$), free energy minimization is necessary, and yields the solution $\mu_1=(\phi_\infty-n\pi)/\log(R/\epsilon_1)$, $\mu_2=0$, $\phi_0=n\pi$, shown in Fig.~\ref{fig:dimer}d-f for different values of $n=-1,0,1$. 
In the limit  $R\to\infty$, the solution is $\mu_1=\mu_2=0$. However, the value of the free energy at the minimum, $F=(\phi_\infty-n\pi)^2/\log(R/\epsilon_1)$, shows that the impact of the far-field boundary condition becomes less and less significant with the system size. With enough space, a very slow spiral with negligible elastic cost is enough to match any enforced far-field orientation $\phi_\infty$. 
If instead, we assume net neutrality from the start, $\mu_1=-\mu_2$, to ensure asymptotically homogeneous far-field, there are enough constraints to compute all three unknowns, and we obtain the solution $\mu_1=-\mu_2=(\phi_\infty-n\pi)/\log(d/\epsilon_1)$, $\phi_0=(\phi_\infty \log (d/\epsilon_2)+n\pi \log(\epsilon_2/\epsilon_1))/\log(d/\epsilon_1)$. Note that this solution does not depend on $R$ at all, because charge neutrality ensures far-field homogeneity which adds no elastic energy when system is enlarged. However, at finite $R$, this solution has a higher free energy than the previously calculated solution which used the extra degree of freedom of nonzero total spiral charge to diminish the elastic energy even further.
 Alternatively, $\phi_\infty$ can be left unconstrained (minimised over), which corresponds to an infinite plane. The way to model infinite systems without imposing an artificial boundary is thus to enforce spiral charge neutrality and minimize over the far-field orientations, in above example, this yields $\phi_\infty =n\pi$. This solution then coincides with the bounded solution in the limit $R\to\infty$.

An important aspect of this problem is that the length scales are hierarchically well separated. The size of the defects must be significantly smaller than the interparticle separation $\epsilon\ll d$, so that the boundary conditions on the circuits of radius $\epsilon$ around the defects can be accurately met. If the particles are too close together, spatial variation of the director field $\phi(z)$ caused by surrounding particles, cause a deviation from the ideal boundary condition, even though the conditions are still matched in an average sense. Such configuration can still be sufficient to create initial conditions for further numerical simulations. A similar condition applies to the size of the entire system that should be much larger than interparticle separation, $R\gg d$. For more tightly confined systems, an exact solution can be obtained if we can find a conformal mapping from the nematic domain to one where boundary conditions can be met, such as used by Tarnavskyy et al.~\cite{TarnavskyyO_LiqCryst45_2018,TarnavskyyOS_LiqCryst47_2020,TarnavskyyOS_LiqCryst}, instead of just using point defects represented by simple zeroes and poles.

An important feature of the energy minimisation described above is that we obtain not only a single solution, but a family of local energy minima, enumerated by $n$, i.e. the number of $\pi$ turns the director makes to match the boundary condition. Each solution does have a different free energy, but the discrete nature of the solutions means they are metastable, and more than one can be realised in experiments, depending on the initial condition. It may also be possible to switch between them with an appropriate manipulation technique.

\section{Many-body generalisation}
\label{sec:manybody}

The minimal model of two interacting defects, described in the previous section, can be generalised to 
a general system of $N$ defects. Using the same procedure of contour integration, we arrive at a general form for the free energy (shown for $\mu $-part, $k$-part being again the same),
\begin{equation}
F_{\mu}=2\pi \left\lbrace \sum_i \mu_i \sum_j \mu_j \log \frac{R}{\epsilon_j}-\sum_{i,j; i\neq j} \mu_i \mu_j\log \frac{d_{ij}}{\epsilon_j}\right\rbrace,
\label{eq:Fsum}
\end{equation}
where $d_{ij}=|z_i-z_j|$ are the distances between the defects. However, we also know that the winding number on the outer boundary must match the sum of all the winding numbers on the inside, but with inverted sense of circulation, so we can assign the outer boundary the charge $\kappa_0=-\sum_i \kappa_i$. This form makes it apparent, that the far boundary acts like just another defect, far enough that distances from all other defects can be treated as equal to $d_{0j}=R$. M\"obius transformations can map any of the defects to infinity while preserving the topology of the solution, so the choice of which defect sits at the infinity and represents the outer boundary, is arbitrary and does not affect the form of the free energy. This is helpful for understanding the symmetry behind the system, but for the further discussion, we will not extend the system of equations with $\kappa_0$. In the limit of a large system size, $R\to\infty$, the first term of Eq.~\ref{eq:Fsum} dominates the second and they decouple in scale. With the diverging logarithmic terms, the first term can only be finite if $\sum_i\mu_i = -\mu_0=0$, and thus the second term is minimised within the states with zero total spiral charge. However, this limit converges extremely slowly, so for simulating unbounded systems, it is beneficial to enforce neutrality exactly, and minimise over the far field boundary orientations,
as discussed in the previous section for the case of a defect pair. We derive the free energy expression for multiple defects in unbounded unconstrained space under assumption of net neutrality in Appendix \ref{appendixB}. The formalism below assumes a fixed far-field at a far boundary at distance $R$ from the defects without the assumption of net-neutrality.

The free energy expression (Eq.~\ref{eq:Fsum}) can be rewritten in the matrix form,
\begin{equation}
F_\mu=2\pi \bm{\mu} M \bm{\mu}=2\pi \bm{\mu}(\Omega - D) \bm{\mu}, \quad \Omega_{ij}=\log\frac{R}{\epsilon_j}, \quad
D_{i\neq j}=\log\frac{d_{ij}}{\epsilon_j}.
\label{eq:Fsummatrix}
\end{equation}
where it is noteworthy that the individual matrices $\Omega$ and $D$ are not symmetric unless the defect cores are all the same size. On the contrary, the introduced energy matrix $M$ that contains the information about the entire system composed of the far-field part $\Omega$ and the pair interaction part $D$, is symmetric and positive semi-definite, as expected for an energy operator (see Appendix~\ref{appendixA} for proof). It can be rewritten, if convenient, as $M_{ij}=\log (R/d_{ij})$, if we define the self-distance $d_{ii}\equiv \epsilon_i$. The matrix for the $k$ part is identical. 

To apply the boundary conditions, we take the same steps we took for the two defects, evaluating the director angle on a path around each defect, and neglecting the displacement from the core radius. Equation \ref{eq:phi_components} rewrites into
\begin{equation}
    \phi(z_i)=k_i \theta_i + \sum_{j\neq i}k_j \theta_{ij}+\phi_0+\sum_{j\neq i}\mu_j \log\frac{d_{ij}}{\epsilon_j}, \quad \theta_{ij}=\arg(z_i-z_j).
\end{equation}
As before, the first term ensures the correct winding number, the second is the contribution of all the rest of the defects where correct branches must be selected for each term, and the last term involves the unknown spiral charges $\mu_j$. The first term by itself can be thought of as the defect in its ``canonical'' orientation (when the director at position $\theta=0$ points along the $x$ axis), the remaining terms modify this orientation due to other defects, and together, they must match the orientation (rotation with respect to the canonical orientation) we enforce at the defect -- the boundary condition $b_i$.

\begin{figure}
    \centering
    \includegraphics[width=\textwidth]{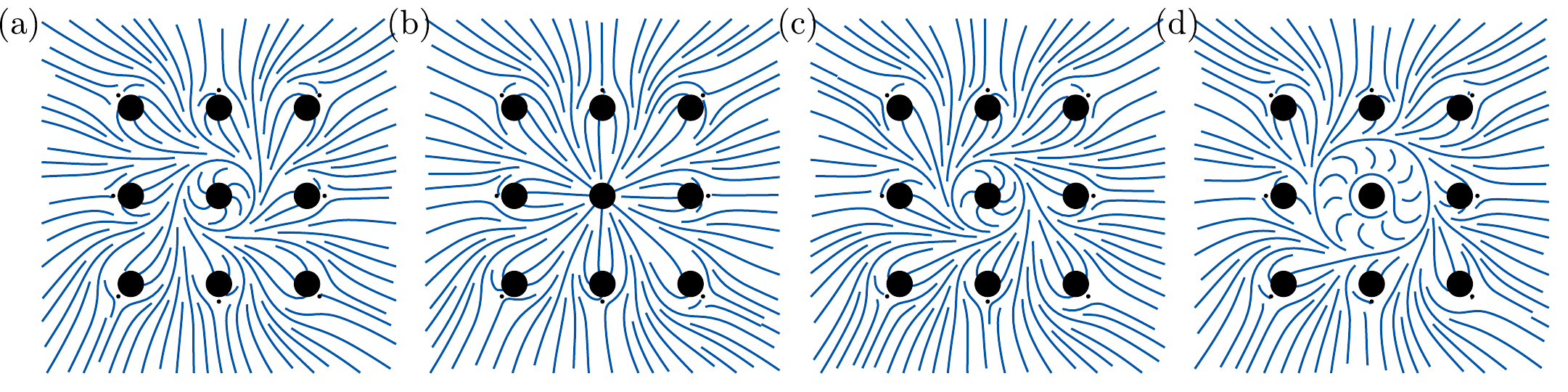}
    \caption{A series of solutions with 9 fixed homeotropic particles, 8 fixed $k=-1$ accompanying defects, for a varying amount extra winding $n\pi$, $n=-1,0,1,2$, for the central particle. Each particle with constrained boundary director has its own discrete index enumerating the local energy minima, although in most cases, extreme spiral charges are not physically plausible.}
    \label{fig:extra-winding}
\end{figure}

The difference $f_i$ between the boundary orientation $b_i$ and the $k$-contribution of the rest of the defects equals the angle that must be accumulated by the $\mu$-part, and may be complemented by an arbitrary multiple of $\pi$ due to rotational symmetries of the director,
\begin{equation}
f_i
\equiv b_i-\sum_{j\neq i} k_j \theta_{ij}+n_i 
\pi
=\phi_0+\sum_{j\neq i} \mu_j \log \frac{d_{ij}}{\epsilon_j}.
\label{eq:boundary_condition_i}
\end{equation}
Figure \ref{fig:extra-winding} shows an example of changing $n_i$ at one of the defects.
The boundary condition on the outer boundary can be written in a similar way,
\begin{equation}
f_0
\equiv b_0+n_0 \pi
=\phi_0+\sum_{j} \mu_j \log \frac{R}{\epsilon_j},
\label{eq:boundary_condition_0}
\end{equation}
where the $k$-contributions of all of the defects in this case simply ensure correct winding.
%We recognize on the left the same matrix elements seen in the free energy. 
Due to the additional degree of freedom $\phi_0$, we can write Eqs.~\ref{eq:boundary_condition_i} and \ref{eq:boundary_condition_0} in the form of an extended $(N+1)\times(N+1)$ linear system:
\begin{equation}
    \begin{bmatrix} f_0 \\ \bm{f}\end{bmatrix} =
    \begin{bmatrix}
    1 & \bm{\omega}^T \\
    \bm{1} & D \\
    \end{bmatrix}\begin{bmatrix}
    \phi_0 \\ \bm{\mu}
    \end{bmatrix},
    \quad \omega_i = \log\frac{R}{\epsilon_i},
    \label{eq:constraintMatrix}
\end{equation}
where $\bm{1}$ is a vector with all values equal to $1$.
The distance matrix $D$ and the matrix $\Omega=\bm{1}\otimes \bm{\omega}$ are the same that appear in the expression for the free energy (Eq.~\ref{eq:Fsummatrix}).

We want to find a free energy minimum corresponding to a constrained orientation of some of the defects. To achieve this, free energy has to be reparameterized in terms of $f_i$ terms rather then $\mu_i$. As some of the defects or the outer boundary might not be necessarily fixed, some of the components $f_i$ may be unconstrained ``slack variables'' that need to be minimized over to find the ground state.
We compute the block-wise inverse of the constraint matrix
\begin{equation}
\begin{bmatrix}
    1 & \bm{\omega}^T \\
    \bm{1} & D \\
    \end{bmatrix}^{-1}=\begin{bmatrix}
    1-\bm{\omega} M^{-1}\bm{1} & \,\bm{\omega} M^{-1} \\
    M^{-1}\bm{1} &  \,-M^{-1} \\
    \end{bmatrix}
    \label{eq:invConstraintMatrix}
\end{equation}
and apply it to Eq.~\ref{eq:constraintMatrix}. The resulting  expression for the spiral charges
\begin{equation}
\bm{\mu}=M^{-1}(f_0 \bm{1}-\bm{f}).
\label{eq:muFromF}
\end{equation}
can be substituted into Eq.~\ref{eq:Fsummatrix}, to obtain the free energy, expressed in terms of $f_i$,
\begin{equation}
    F=2\pi \bm{k}M\bm{k}+2\pi\begin{bmatrix} f_0 \\ \bm{f}\end{bmatrix}^T 
    \begin{bmatrix}
        \bm{1}M^{-1}\bm{1} & -\bm{1}M^{-1} \\
        -M^{-1}\bm{1} & M^{-1} \\
    \end{bmatrix}
    \begin{bmatrix} f_0 \\ \bm{f}\end{bmatrix},
    \label{eq:energy_final}
\end{equation}
which as a quadratic functional can easily be minimised with respect to any subset of $f_i$ that are not yet constrained by the boundary condition. This procedure can solve for any number of constraints, from a fully constrained case, where inversion of Eq.~\ref{eq:constraintMatrix} is the only step, and no minimisable degrees of freedom remain, to the fully unconstrained case with the trivial solution $\mu_i=0$.

For any constraint, there is an infinite discrete set of local minima, parameterized by indices $n_i$, as shown for selected values of $n_i$ in Fig.~\ref{fig:extra-winding}. The meaning of these indices reveals that a system of defects with $m$ constraints has $m-1$ integer-valued topological indices (excluding one due to the fact that adding the same multiple of $\pi$ to the entire director changes nothing). Varying the index $n_i$ changes the number of relative half-turns the director makes between $i$-th defect and all other \emph{constrained} defects. A system of a large number of constrained 2D defects, such as a lattice of micropillars, colloidal inclusions or other topological obstructions with fixed boundary conditions, will thus always be multistable, with a multidimensional lattice of multistable states. Although there will be a practical limit on heavily wound spirals that are hard to create and stabilize, we expect at least a limited number of metastable states. Although this solution was derived for small circular defect-inducing boundaries, the topology and multiplicity of solutions will hold for inclusions of arbitrary shape with topology of a disk, as long as no additional singularities appear on the surface itself.

The values of the indices $n_i$ can be set in advance to obtain solutions for each set of indices. Alternatively, we can view the final system of equations as a quadratic minimization problem with $m-1$ discrete variables, $N-m$ continuous variables and one zero-mode. In this view, constraining the director on each boundary quantizes one dimension of the parameter space.

\section{Forces and torques}

The gradient of the free energy (Eq.~\ref{eq:energy_final}) with respect to individual defect coordinates $z_i$ gives a quasi-static approximation of a force acting on each defect, assuming the director reconfiguration time is much faster than the defect motion. 
The dependence on inter-particle distances resides both in the matrix $M$ (defined in Eq.~\ref{eq:Fsummatrix}) and in $\bm{f}$ terms (defined in Eq.~\ref{eq:boundary_condition_i}) for defects with prescribed boundary conditions. The gradient can thus be symbolically (keeping in mind matrix and vector nature of the variables) written as
\begin{equation}
    \nabla_i F = \frac{\partial F}{\partial M}\nabla_i M
    +\frac{\partial F}{\partial \bm{f}} \nabla_i \bm{f},
\end{equation}
with $\nabla_i$ representing the derivative with respect to the coordinates $z_i$ of $i$-th defect. The first term reduces to $2\pi\bm{k}\nabla_i M \bm{k}-2\pi \bm{\mu}\nabla_i M \bm{\mu}$, with the minus sign stemming from the fact, that the free energy is differentiated at constant $\bm{f}$, not at constant $\bm{\mu}$.
The second term is zero for unconstrained degrees of freedom, which are minimised over $\bm{f}$.
%due to the $k$-part of Eq.~\ref{eq:boundary_condition_i} containing dependence on the relative angles $\theta_{ij}$ with respect to the defects. 
After some simplification, we obtain the force on the i-th defect
\begin{equation}
\mathcal{F}_i=-\nabla_i F=-2\pi \left[\bm{k}\nabla_i{M}\bm{k}-\bm{\mu}\nabla_i{M}\bm{\mu}-2\bm{\mu}\nabla_i \bm{f}\right],
\label{eq:force}
\end{equation}
where $\bm{f}$ was expressed in terms of $\bm{\mu}$ to simplify the expression. The spiral charges $\bm{\mu}$ are the solution of the minimization Eq.~\ref{eq:energy_final} and depend on which constraints are imposed.

The first term gives rise to pairwise central forces obeying an inverse distance law:
\begin{equation}
    2\pi(\bm{k} \nabla_i M \bm{k})=-4\pi \sum_{j; j\neq i}k_ik_j\frac{z_i-z_j}{d^2_{ij}},
\end{equation}
The second term has the same direction and distance scaling for the spiral charge contributions.
The final term contains the effect of the boundary condition vector $\bm{f}$. The components with no constraints (the slack variables) do not contribute to this term, while the constrained components differentiate as
\begin{equation}
    \nabla_i f_j = -\sum_{l\neq j} k_l \nabla_i \theta_{jl}=-\sum_{l\neq j} k_l\frac{\ii (z_i-z_l)}{d_{il}^2}\delta_{ij}
    -k_i \frac{\ii (z_i-z_j)}{d_{ij}^2}(1-\delta_{ij}),
    \label{eq:gradf}
\end{equation}
giving a contribution of constraint $j$ to the force on the particle $i$. This term is weighted by the coefficients $\bm{\mu}$, so the amount of spiral charge regulates the amount each defect can influence others. The presence of $\ii$ multiplied by the distance vector between defect pairs, describes a tangential force that is perpendicular to the distance vector.
The first term in Eq.~\ref{eq:gradf} represents the reaction force acted upon a constrained particle itself by all other particles, while the second term in Eq.~\ref{eq:gradf} is the force exerted by a constraint on a different particle.

\begin{figure}
    \centering
    \includegraphics[width=0.66\textwidth]{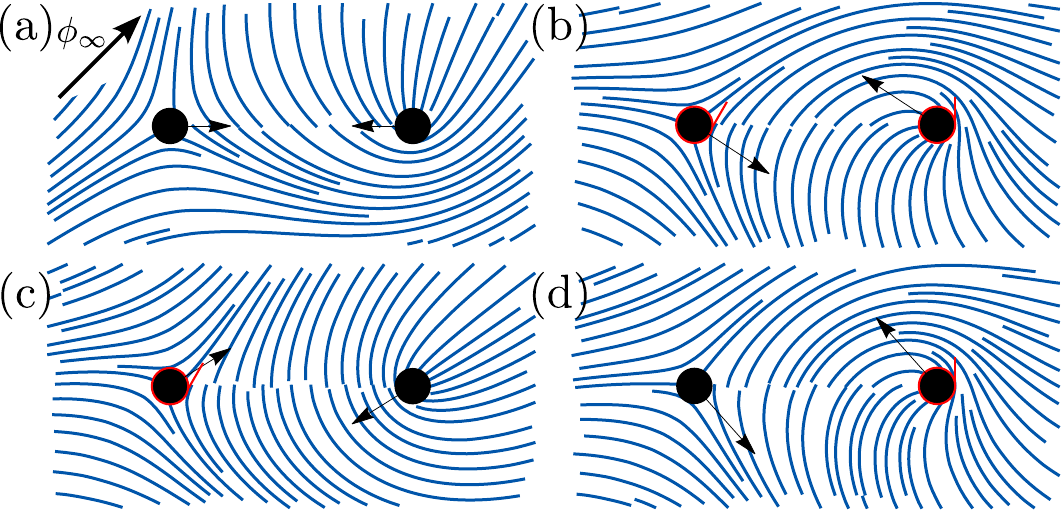}
    \caption{Forces acting on a $\pm \frac12$ defect pair with different constraints. (a) Without constraints, the forces are radial. (b-d) Defects with fixed boundary condition are circled in red, with a red line indicating the preferred direction at $\theta_i=0$ position. Presence of boundary conditions induces tangential force components. The far field direction, marked in the corner, is fixed at the distance $R=10$, compared to the inter-particle distance $d_{12}=2$ and defect radii $\epsilon=0.15$.
    }
    \label{fig:forces}
\end{figure}

\begin{figure}
    \centering
    \includegraphics[width=0.66\textwidth]{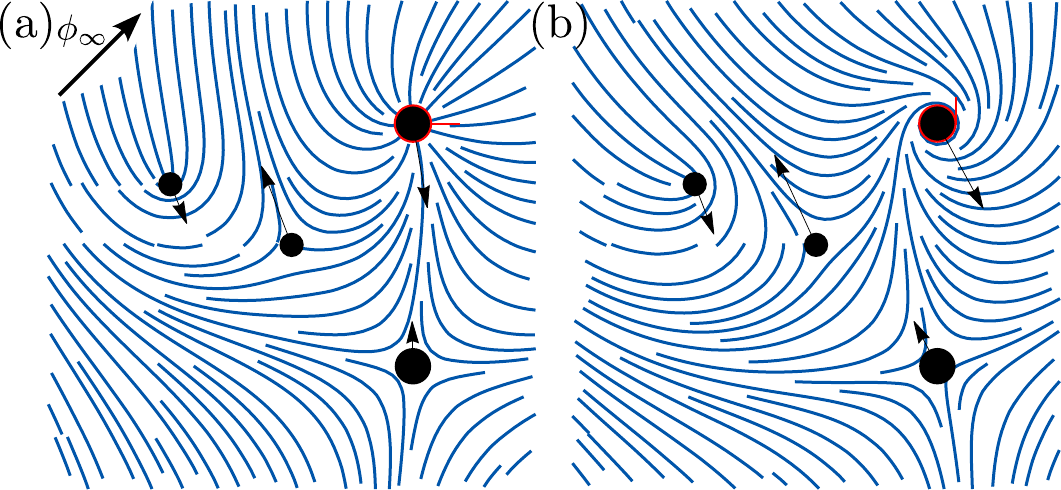}
    \caption{Forces acting on a a set of $\pm \frac12$ and $\pm 1$ defects, with a boundary condition on a single $+1$ defect, which could represent a micropillar or a colloidal inclusion. Homeotropic and planar alignment results in different forces.
    The boundary condition is at $R=10$, the size is $\epsilon=0.15$ for the large defects and $\epsilon=0.1$ for the small defects.
    }
    \label{fig:forces4}
\end{figure}

We illustrate the importance of orientational constraints on the forces acting on the defect pair in Fig.~\ref{fig:forces}. Without orientational contraints, all spiral charges are zero and the forces are central (Fig.~\ref{fig:forces}a). The magnitude and the direction of forces changes if both defects are constrained (Fig.~\ref{fig:forces}b) or if one of them is constrained (Fig.~\ref{fig:forces}c,d). 
One possibility to prescribe the orientation of a defect is to determine the anchoring of nematic molecules at the surface of a colloidal inclusion with winding number $k=+1$. In Figure~\ref{fig:forces4}, we show how switching between homeotropic and tangential anchoring at the surface of one inclusion alters the forces on each defect within the system.

In addition to forces, the director field also imparts torques on the particles. Torques reflect energy minimization by changing the boundary condition vector $\bm{b}$ in Eq.~\ref{eq:boundary_condition_i}. However, rotation of the boundary condition is nontrivially connected to rotation of particles. For instance, a particle with $k=1$ is invariant to rotation, and for a particle with $k=0$, rotation of the particle by some angle results in rotation of the boundary condition by the same angle. 
Considering the free energy expression in Eq.~\ref{eq:energy_final},the torque on  $i$-th particle reduces to
\begin{equation}
    \mathcal{T}_i=-(1-k_i)\frac{\partial F}{\partial b_i}=(1-k_i)4\pi \mu_i.
\end{equation}
The spiral charge $\mu_i$ therefore has a direct interpretation as a restoring torque that aims to unwind the spiral.

\begin{figure}[h]
    \centering
    \includegraphics[width=\textwidth]{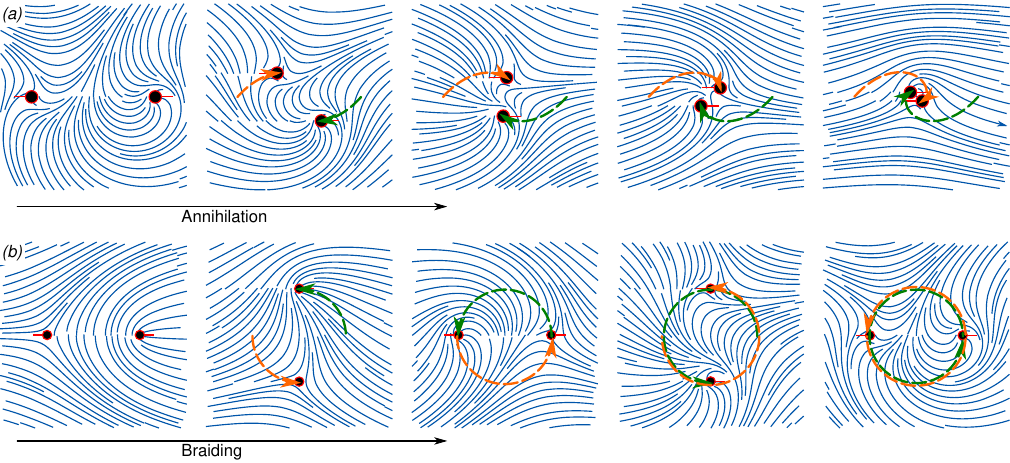}
    \caption{Annihilation and braiding of a $\pm1/2$ defect pair with a fixed orientation. Defects have equal core sizes ($\epsilon_1=\epsilon_2=\epsilon$). 
    (a) Non-central forces lead to spiral annihilation trajectories. The initial separation is set to $d_{12}=20\epsilon$.
    (b) Defects with the same separation of $d_{12}=20\epsilon$ are initiated without any spiral charge ($\mu_1=\mu_2=0$). By braiding the defect pair with a fixed orientation in the counter-clockwise direction around each-other, the $+1/2$ defect obtains a spiral charge of $-\pi/\log(20)$ and the $-1/2$ defect obtains a spiral charge of  $\pi/\log(20)$, in line with Eq.~\ref{eq:braiding}.
    }
    \label{fig:brading}
\end{figure}

With forces and torques computable for every configuration, one can write down effective dynamic equations
\begin{equation}
    \dot{z}_i=\gamma_{ti}\mathcal{F}_i ,\quad \dot{\beta}_i=\gamma_{ri} \mathcal{T}_i,
\end{equation}
where $\gamma_{ti}$ and $\gamma_{ri}$ are the translational and rotational drag coefficients of the $i$-th particle, respectively, and $\beta_i=b_i (1-k_i)^{-1}$ is the physical rotation angle of the particle.
Fig.~\ref{fig:brading}(a) shows an example, how the constraints on the defect orientations lead to spiral trajectories in the case where a $\pm 1/2$ defect pair is let to annihilate at fixed defect orientations.
Without the orientational constraints, the defects would annihilate in a straight line. Using finite translational and orientational drag coefficients, one could in principle observe a range of annihilation trajectories from straight lines to spirals.

Spiral charge can be modified by not only changing the orientations of defects, but also by changing defect positions.  Eq.~\ref{eq:boundary_condition_i} tells us that the orientation of a defect is determined from the contributions from all other defects through the boundary condition $b_i$: 
\begin{equation}
b_i=\phi_0+\sum_{j} \mu_j \log \frac{d_{ij}}{\epsilon_j}+\sum_{j\neq i} k_j \theta_{ij}+n_i \pi,
\label{eq:orientation}
\end{equation}
where $n_i$ is an integer. For a fixed defect orientation (at constant $b_i$), a small displacement of defect positions results in the necessary change of the spiral charges $\mu_i$. By braiding the $i$-th defect with a fixed orientation around the $j$-th defect for a full counterclockwise turn around each other, the spiral charges of the two defect change by
\begin{equation}
    \Delta\mu_i=-\frac{2\pi k_i}{\log\frac{d_{ij}}{\epsilon_i}}.
    \label{eq:braiding}
\end{equation}
We demonstrate such change of the spiral charges in Fig.~\ref{fig:brading}(b) by braiding a $\pm1/2$ defect pair for a full loop around each other, while keeping the orientations of both defects fixed.

\section{Discussion}

The presented complex field approach to construction of harmonic solutions in a planar orientational field with polar, nematic, or higher order symmetry, is developed to solve the field around sets of point defect structures or small enough circular inclusions or boundaries. The solutions allow for an arbitrary number and orientation of disclinations and are represented as a combination of topological charge (winding number $k$) and spiral charge $\mu$. As shown by Pearce and Kruse, without spiral contributions the generated solutions can not be mapped to director fields observed in experiments~\cite{PearceDJG_SoftMatter17_2021}. The benefit of our framework is that the generated fields solve the Laplace equation and correspond to a free energy minimum with a single elastic constant. The forces and torques on the defect structures can also be analytically expressed. Non-central forces and non-zero torques are a direct consequence of the spiral charge. For the simplest configuration of only two defects, the presence of the spiral charge leads to spiral annihilation trajectories in agreement with the literature~\cite{TangX_SoftMatter13_2017,PearceDJG_SoftMatter17_2021}. Although at given positions and orientations of defects, the director field is in equilibrium, the computed forces and alignment torques are still relevant for defect dynamics in active and non-equilibrium liquid crystals~\cite{shankar2018defect,kumar2018tunable,thijssen2020role,zhang2018interplay}.
The defects act as quasi-particles under the effect of pair interactions, and allow planar nematic simulations at the level of dissipative particle dynamics without the need to model the nematic host. As the spiral charges change in response to motion due to to boundary conditions, their motion is collective, and not reducible to a simple pair potential, but the resulting system is linear, and as such is trivial to compute numerically for a moderate number of particles.

We show that movement of defects with a fixed orientation is directly linked to changes of the spiral charge, similarly to how anyonic braiding was shown in $k$-atic liquid crystals by modulation of the boundary condition~\cite{mietke2022anyonic}. Braiding of defects was observed also for spiral waves in living cells~\cite{liu2021topological} and in active nematics it was associated with topological entropy and chaos~\cite{tan2019topological,smith2022braiding,mitchell2024maximally}.
Harmonic director fields are of great interest also in three dimensions~\cite{binysh2018maxwell}. While our theory considers purely two-dimensional fields, a future challenge would be to address the defect textures also in quasi-two-dimensional layers, where the director is allowed to point out of plane.

Construction of director structures can be extended from simple rational expressions describing finite sets of defects, to any analytical functions with appropriate positioning of poles and zeros to account for infinite sets of defects, such as periodic defect arrays. The formalism remains the same, but the tractability of the elastic energy integrals depends on the function. For example, linear chains of dipoles can be mapped to trigonometric functions while doubly periodic lattices involve Jacobi or Weierstrass elliptic functions.

In this manuscript, we only considered point defect \emph{monopoles}, which only includes complexified angle functions $\Phi$ with logarithmic singularities. Our formalism can be extended to include higher multipoles with singularities of the form $z^{-\ell}$ for $\ell$-th multipole (see \cite{houston2023colloids}), which do not require addition of spiral contributions. This would allow to account for particles with more elaborate director profiles, described through a multipolar expansion, at the expense of more convoluted expressions in the interaction matrix $M$ and boundary constraints.

Our construction is not directly meant to solve the director field on domains that involve nontrivial boundary shapes or larger inclusions. For such problems, see other works that employ conformal mapping to transform the domain to one that offers a solution in terms of simple analytical functions --- such solutions are limited to geometries that have an analytically tractable conformal mapping \cite{DavidsonAJ_EurJApplMath23_2012,TarnavskyyOS_LiqCryst47_2020,TarnavskyyOS_LiqCryst,vafa2023periodic}, or can otherwise be solved numerically \cite{vanBijnenRMW_PhysRevE86_2012}. Our approach is a good approximation for most systems of small, preferably spherical particles. What is lost in exactness of boundary conditions, is offset by the fact that a closed-form solution can be obtained by a routine linear algebra algorithm for an arbitrary number of constrained boundaries. Here, we assume an infinitely strong anchoring, with director at the boundary prescribed exactly. For handling of finite anchoring at arbitrarily shaped boundaries, see Chandler and Spagnolie \cite{chandler2023nematic}.
Finally, the method's ability to handle larger numbers of defects suggests a possibility of developing a numerical scheme capable of handling finite-sized boundaries of arbitrary shapes, akin to methods used in hydrodynamics and electrostatics. We leave this as a future challenge.

\begin{acknowledgments}
The authors acknowledge funding from Slovenian Research and Innovation Agency (ARIS) under contracts P1-0099, J1-50006, and N1-0195.
\end{acknowledgments}

\appendix

\section{Positive semi-definiteness of free energy}
\label{appendixA}
We show that the free energy expression
\begin{equation}
F=2\pi \left\lbrace \sum_i \kappa_i^\ast \sum_j \kappa_j \log \frac{R}{\epsilon}-\sum_{i\neq j} \kappa_i^\ast \kappa_j\log \frac{d_{ij}}{\epsilon}\right\rbrace,
\label{eq:F_positive}
\end{equation}
obtained from Eq.~\ref{eq:Fsum} for the defect cores of the same size, is always non-negative, provided that the total charge of all N defects sums up to zero:
\begin{equation}
\sum_{i=1}^N \kappa_i=0
\end{equation}
and that all defects are at least $2\epsilon$ apart from each other ($d_{ij}\geq2\epsilon$ for $i\neq j$). 

The first term in Eq.~\ref{eq:F_positive} is directly set to zero once the zero total charge condition is applied. We can rewrite the second term of the free energy expression
\begin{equation}
F=-2\pi \sum_{i=2}^N \left[ \left( {\kappa}_1^*{\kappa}_i + {\kappa}_1{\kappa}_i^*\right)\log\frac{d_{1i}}{\epsilon} + \sum_{j=2, j\neq i}^N  {\kappa}_i^*{\kappa}_j \log\frac{d_{ij}}{\epsilon} \right]
\end{equation}
and set ${\kappa}_1=-\sum_{i=2}^N{\kappa}_i$, obtaining
\begin{equation}
F=-2\pi \sum_{i=2}^N \left[ -\sum_{j=2}^N \left( {\kappa}_i{\kappa}_j^* + {\kappa}_i^*{\kappa}_j \right)\log\frac{d_{1i}}{\epsilon} + \sum_{j=2, j\neq i}^N  {\kappa}_i^*{\kappa}_j \log\frac{d_{ij}}{\epsilon} \right],
\end{equation}
which can be rewritten in a matrix form as
\begin{equation}
F=2\pi\tilde{\bm{\kappa}}^* \tilde{D} \tilde{\bm{\kappa}},
\end{equation}
where $\tilde{\bm{\kappa}}=[\kappa_2, \dots, \kappa_N]^\intercal$ and
\begin{equation}
\tilde{D}_{ii}=2\log\frac{d_{1i}}{\epsilon},\quad \tilde{D}_{i\neq j}=-\log\frac{d_{ij}}{\epsilon}+\log\frac{d_{1i}}{\epsilon}+\log\frac{d_{1j}}{\epsilon}
\end{equation}
with $i,j\in\{2,\dots ,N\}$,
\begin{equation}
\tilde{D}=\begin{pmatrix} 
2\log\frac{d_{12}}{\epsilon} & \log\frac{d_{12}}{\epsilon}  + \log\frac{d_{13}}{\epsilon}  - \log\frac{d_{23}}{\epsilon}\quad  & \cdots & \log\frac{d_{12}}{\epsilon}  + \log\frac{d_{1N}}{\epsilon}  - \log\frac{d_{2N}}{\epsilon}  \\
\log\frac{d_{12}}{\epsilon}  + \log\frac{d_{13}}{\epsilon}  - \log\frac{d_{23}}{\epsilon}\quad  & 2\log\frac{d_{13}}{\epsilon}  & \cdots & \log\frac{d_{13}}{\epsilon}  + \log\frac{d_{1N}}{\epsilon}  - \log\frac{d_{3N}}{\epsilon}  \\
\vdots\quad & \vdots & \ddots & \vdots \\
\log\frac{d_{12}}{\epsilon}  + \log\frac{d_{1N}}{\epsilon}  - \log\frac{d_{2N}}{\epsilon}\quad  & \log\frac{d_{13}}{\epsilon}  + \log\frac{d_{1N}}{\epsilon}  - \log\frac{d_{3N}}{\epsilon}  & \cdots & 2\log\frac{d_{1N}}{\epsilon} 
\end{pmatrix}
\end{equation}
The free energy is positive for arbitrary $\tilde{\bm{\kappa}}$, as long as the real symmetric matrix $\tilde{D}$ is positive-definite.
We can use the Sylvester's criterion for positive semi-definiteness, which states that $\tilde{D}$ is positive semi-definite, if and only if all principal minors of $\tilde{D}$ are non-negative.

For $k=N-1$, we can construct a simplex with $N$ vertices in a k-dimensional space, where the Euclidean distances between vertices correspond to $L_{ij}^2=\log\frac{d_{ij}}{\epsilon}$. Note that this is possible because the condition $d_{ij}\geq2\epsilon$ for $i\neq j$ leads to non-negative values of $\log\frac{d_{ij}}{\epsilon}$ and also to the triangle inequality $\log\frac{d_{ij}}{\epsilon}+\log\frac{d_{jk}}{\epsilon}\geq \log\frac{d_{ik}}{\epsilon}$. The k-dimensional volume of the simplex is always non-negative and is given by the Cayley–Menger determinant~\cite{Sommerville}
\begin{align}
V^2 & = \frac{1}{(k!)^2 \, 2^k}
\begin{vmatrix} 
2L_{12}^2 \quad & L_{12}^2 + L_{13}^2 - L_{23}^2 & \cdots & L_{12}^2 + L_{1N}^2 - L_{2N}^2 \\
L_{12}^2 + L_{13}^2 - L_{23}^2 \quad & 2L_{13}^2 & \cdots & L_{13}^2 + L_{1N}^2 - L_{3n}^2 \\
\vdots \quad & \vdots & \ddots & \vdots \\
L_{12}^2 + L_{1N}^2 - L_{2N}^2 \quad & L_{13}^2 + L_{1N}^2 - L_{3N}^2 & \cdots & 2L_{1N}^2
\end{vmatrix}
\geq 0.
\end{align}
Using the expression for the simplex volume, we have shown that $\det(\tilde{D})=(k!)^2 \, 2^k V^2\geq 0$. All other principal minors of $\tilde{D}$ effectively correspond to the same calculation for a smaller number of defects and can be shown to be non-negative using the same arguments. Following the Sylvester's criterion, we can therefore conclude that the free energy is indeed positive semi-definite ($F\geq 0$) for an arbitrary number of defects, provided that total charge of defects is zero and defects are sufficiently spaced from each other.

\section{Solution for an unbounded system}
\label{appendixB}

In section \ref{sec:manybody}, we provided a solution that allows solving for spiral charges in a finite system bound into a circular domain of radius $R$, with constraints given at any combination of boundary conditions, including at the outer boundary. This requires a mapping between unknown quantities $\phi_0, \bm{\mu}$ and quantities $f_0, \bm{f}$, which can either be given by boundary conditions or let to vary to minimise the free energy.

An unbounded system does not have an outer boundary and thus also has no boundary condition given there. Instead, the sum of spiral charges must equal zero. To solve for this, we must make $f_0$ an unknown quantity, specifying the far-field orientation, instead of the overall rotation $\phi_0$, which can be eliminated as an unknown and can be computed from Eq.~\ref{eq:boundary_condition_0} after the calculation.
Using Equation \ref{eq:muFromF} together with the zero total charge condition, which can be written as $\bm{1}\cdot \bm{\mu}=0$, we can write a modified version of Eq.~\ref{eq:constraintMatrix}:
\begin{equation}
    \begin{bmatrix} 0 \\ \bm{f}\end{bmatrix} =
    \begin{bmatrix}
    0 & \bm{1}^T \\
    \bm{1} & -M \\
    \end{bmatrix}\begin{bmatrix}
    f_0 \\ \bm{\mu}
    \end{bmatrix}.
    \label{eq:constraintMatrixNeutral}
\end{equation}
This matrix does not depend on $R$, as we eliminated the outer boundary.

Repeating the block-wise inverse Eq.~\ref{eq:invConstraintMatrix}, we obtain,
\begin{equation}
    \begin{bmatrix}
    0 & \bm{1}^T \\
    \bm{1} & -M \\
    \end{bmatrix}^{-1}
    =\begin{bmatrix}
    r & r\bm{1}M^{-1} \\
    r M^{-1}\bm{1} & r(M^{-1}\bm{1}\otimes \bm{1}M^{-1})-M^{-1} \\
    \end{bmatrix},\quad r = (\bm{1}M^{-1}\bm{1})^{-1},
\end{equation}
and reformulate the free energy in the form,
\begin{equation}
    F=2\pi\bm{k}M\bm{k} + 2\pi \bm{f}M^{-1}\bm{f}-2\pi r(\bm{f}M^{-1}\bm{1})^2.
    \label{eq:FNeutral}
\end{equation}
Dependence of the matrix $M$ and the fixed boundary conditions $\bm{f}$ on defect positions remain unchanged, so their derivatives are unaffected by the changed boundary condition at infinity.
The expression for the force (Eq.~\ref{eq:force}) remains unchanged, but with different values of $\bm{\mu}$, obtained from the solution of Eq.~\ref{eq:constraintMatrixNeutral} and minimization of Eq.~\ref{eq:FNeutral}.

\bibliography{references}

\end{document}